\documentstyle[aps,preprint]{revtex}
\begin{document}
\draft
\sloppy
\newcommand{\beq}{\begin{equation}}
\newcommand{\eeq}{\end{equation}}
\newcommand{\bea}{\begin{eqnarray}}
\newcommand{\eea}{\end{eqnarray}}
\newcommand{\s}{\sum_{i}}
\newcommand{\su}{\sum_{i<j}}
\newcommand{\wpi}{\sqrt{\hat{p}_{i}^{2}+m^{2}}}
\newcommand{\wpj}{\sqrt{\hat{p}_{j}^{2}+m^{2}}}
\newcommand{\wpa}{\sqrt{\hat{p}_{1}^{2}+m^{2}}}
\newcommand{\wpb}{\sqrt{\hat{p}_{2}^{2}+m^{2}}}
\newcommand{\ep}{\epsilon(\hat{x}_{1}-\hat{x}_{2})}
\newcommand{\xd}{(\hat{x}_{1}-\hat{x}_{2})}
\newcommand{\h}{\hat{H}}
\newcommand{\p}{\hat{P}}
\newcommand{\k}{\hat{K}}
\newcommand{\q}{\hat{Q}}
\newcommand{\pr}{\prod_{k} \theta(\hat{p}_{k})}
\newcommand{\prv}{\prod_{k,k\neq i}\theta(\hat{p}_{k})}
\newcommand{\xa}{\hat{x}_{1}}
\newcommand{\xb}{\hat{x}_{2}}
\newcommand{\ka}{\hat{\Omega}_{1}}
\newcommand{\kb}{\hat{\Omega}_{2}}
\newcommand{\ia}{\hat{p}_{1}}
\newcommand{\ib}{\hat{p}_{2}}

\title{Example of a Poincar\'e anomaly  in relativistic
quantum mechanics\thanks{This work is supported in part by funds provided
by the U. S. Department of Energy (D.O.E.) under cooperative research
agreement DE-FC02-94ER40818 (B.S.) and DE-FG02-91ER40608 (S.L.)
and by a Feodor-Lynen-Fellowship of the
Alexander von Humboldt-Foundation (B.S.).}
}

\author{Stefan Lenz}

\address{Center for Theoretical Physics, Sloane Physics Laboratory\\
         Yale University, New Haven, CT 06520-8120, U.S.A.}

\author{Bernd Schreiber}

\address{ Massachusetts Institute of Technology\\
Center for Theoretical Physics, Cambridge, MA 02139, U.S.A.  }

\date{\today}
\maketitle
\begin{abstract}
The Poincar\'e algebra of classical electrodynamics in
one spatial dimension is studied using light-cone coordinates  and
ordinary Minkowski coordinates. We show that it is possible
to quantize the theory by a canonical quantization
procedure in a Poincar\'e invariant manner
on the light-cone. We also show that this is not
possible when using ordinary coordinates. The physical
reason of this anomaly is analysed.
\end{abstract}

\pacs{PACS number: 03.65.Pm, 11.10.Ek, 11.10.Kk}

\newpage

\section{Introduction}

It is well known that it is generally not possible to
formulate relativistic quantum mechanics with a
finite number of degrees of freedom, due to 
the existence of negative energy solutions of the relativistic
wave equations. These solutions are interpreted as
antiparticles. Wave packets with an extension smaller
than the particle's Compton wavelength\footnote{
Throughout this paper we set $\hbar = c = 1$.
}, $\lambda_{c} = 1/m$,
contain considerable admixtures of these antiparticle solutions
(see e.g. \cite{Fesh}).
Therefore a many body description of relativistic quantum
systems is usually inevitable.

We will  show, however, that it is possible to quantize
gauge theories in one space and one time  dimension (1+1-dimensions)
on a truncated Fock space, while preserving Lorentz invariance, if one
uses light-cone coordinates. More specifically,
in this  paper we will analyse the difference between light-cone
quantization and quantization in ordinary coordinates in
a Hamiltonian formulation of electrodynamics in 1+1-dimensions.
We will explicitly show that canonical quantization of this
theory is possible in a Poincar\'e invariant manner
in light-cone coordinates but not
in ordinary coordinates.
This result means that one encounters a quantum mechanical
 anomaly when using an equal-time
quantization procedure.
 Position space matrix elements
for  the anomalous contribution to the Poincar\'e algebra
can be interpreted in terms of
particle trajectories which make an intuitive interpretation
of the results possible.

In a pioneering paper, Dirac formulated the basic requirements
necessary for combining the Hamiltonian formulation of dynamics with
special relativity \cite{Dirac}. The principle of relativity demands
invariance of physical laws under continuous coordinate transformations,
which are boosts and translations\footnote{
  In 3+1 dimensions also invariance under rotations has to be demanded.
}
\beq
x^{\mu'}=\Lambda_{\ \nu}^{\mu}x^{\nu}+a^{\mu} \qquad
\Lambda_{\mu}^{\ \nu}\Lambda^{\mu\rho}=g^{\nu\rho}.
\eeq
This so called Poincar\'e group is determined by the
Lie algebra of its generators
\begin{equation}
   \{ H , P \} = 0, \quad
   \{ K , P \} = H, \quad
   \{ K , H \} = P.
   \label{Poinc}
\end{equation}
$H := P^0 $ is the Hamiltonian, $P := P^1$ is the momentum and $K$ the
boost generator of the physical system under consideration.
$\{A,B\}$ is the standard
Poisson bracket \cite{Goldstein}, with $\{x_n,p_m \} = \delta_{nm}$.
$x_m $ are the positions of the particles and $p_m $ the
corresponding canonical momenta.
Note that although a Hamiltonian formulation
is not manifestly covariant, the theory is still Lorentz invariant
if the Poincar\'e algebra is closed.
$K$ generates a canonical transformation
of the phase space variables $(x_m , p_m )$
which connects different reference frames.
Due to the last two brackets in (\ref{Poinc}),
$P^\mu = (P^0 , P^1 )$ transforms like a Lorentz vector.
The first bracket expresses translation invariance.

Quantum mechanically, one has to replace
the Poisson brackets by $-i[\ ,\ ]$ where $[\ ,\ ]$ is the
commutator.
The quantum Poincar\'e algebra therefore is
\beq
   [ \hat{H} , \hat{P} ] = 0 , \quad [ \hat{K} , \hat{P} ] = i\hat{H},
   \quad [ \hat{K} , \hat{H} ] = i \hat{P}
   \label{Qpoinc}
.
\eeq
A closed quantum Poincar\'e algebra means that
a unitary transformation
\[
  \hat{U} (\lambda ) := \exp ( i \lambda \hat{K} )
\]
exists
which connects quantities in different Lorentz frames, where
$\hat K$ is the generator of this unitary transformation.
The existence of this unitary transformation is a necessary
and sufficient condition for a  Lorentz invariant formulation
of quantum mechanics.
In general the existence of a canonical transformation does
not guarantee the existence of the corresponding unitary
transformation. One example for this is the axial anomaly
in QED where the classical theory is invariant under chiral
rotations but the quantized theory is not \cite{Adler,Bell,Fuji}.

A Hamiltonian formulation of a dynamical system using
ordinary Minkowski coordinates $x^0 ,x^1$ corresponds
to specifying the initial conditions on a space-like
hypersurface $x^0 = {\rm const}$  in Minkowski space.
Another possibility is offered by the use of light-cone coordinates
\beq
   x^{+} = \frac{x^0 + x^1 }{\sqrt{2}} \qquad
   x^- = \frac{x^0 - x^1}{\sqrt{2}}
   \label{lcc}
\eeq
where a new time variable $x^{+}$ and a new space
variable $x^{-}$ are introduced \cite{Dirac}.
Initial conditions are now specified on a
lightlike hypersurface $x^+ = {\rm const}$.
As with the ordinary coordinates, the generators of the Poincar\'e group are
the components of the momentum vector:
\beq
  H_{LC} := P_{+} = \frac{P_0 + P_1}{\sqrt{2}} \quad
  P_{LC} := P_{-} = \frac{P_0 - P_1}{\sqrt{2}}
  \label{Ppm}
\eeq
and a boost generator $K_{LC}$.
The quantum Poincar\'e algebra has the form
\beq
   [ \hat{H}_{LC} , \hat{P}_{LC} ] = 0 \quad
   [ \hat{K}_{LC} , \hat{H}_{LC} ] = - i \hat{H}_{LC} \quad
   [ \hat{K}_{LC} , \hat{P}_{LC} ] =  i \hat{P}_{LC}
   \label{Poinc2}
   .
\eeq
These relations are found by using the definitions of
$P_+$ and $P_-$ (\ref{Ppm}) and the Heisenberg brackets (\ref{Qpoinc}).
Quantum mechanically, the choice of one of the two formulations
does not necessarily lead to the same theory. This will be demonstrated for
classical electrodynamics in one space and one time dimension below.

This paper is organized as follows: In section \ref{Class}
we will show that the Poincar\'e algebra of classical electrodynamics
closes in 1+1-dimensions. The quantum Poincar\'e algebra, however,
does not close
when one is using ordinary coordinates.
The physical reason for this anomaly will be explained in
terms of retardation effects.
Analogous calculations using light-cone coordinates show that
no anomaly occurs  in this case (sec. \ref{Light}).
In section \ref{Infinit} we show that the anomalous
contribution to the Poincar\'e algebra vanishes in
the infinite momentum frame. We conclude in section \ref{ENDE}.

\section{Poincar\'e invariance for ordinary coordinates}
\label{Class}

In one space and one time dimension (1+1 dimensions)
the field strength tensor contains only one independent component, and
the electric field can easily be expressed by the coordinates of the charged
particles \cite{Bialynicki}. In this paper, we restrict ourselves
to 2 particles of equal mass m with  charges $e$ and $-e$. 
The generalization to many particles
is straightforward.
One obtains the following classical Hamilton
function
\begin{equation}
 H = \sum_{i=1}^{2} \sqrt{p_{i}^{2} + m^{2}} + e^2  |x_{1} -x_{2}|.
\label{xy}
\end{equation}
The remaining two generators of the Poincar\'e algebra are
\begin{eqnarray}
 P &  = &  \sum_{i=1}^{2} p_{i} \\
 K & = &   \frac{1}{2} \sum_{i=1}^{2}
             ( x_{i} \sqrt{p_{i}^{2} + m^{2}} +
             \sqrt{p_{i}^{2} + m^{2}} x_i )
            +
      \frac{1}{2} e^2
      \epsilon (x_{1} - x_{2} ) (x_{1}^{2} - x_{2}^{2} )
\label{xyy}
\end{eqnarray}
($\epsilon (x)$ is 1 for positive and -1 for negative argument).
It is a simple exercise to show that the Poincar\'e algebra (\ref{Poinc})
is closed. The ordering of the first two terms in $K$ is of no significance
in the classical theory, but is needed in order to ensure hermiticity
of $K$ after quantization.

The reason why it is possible to formulate electrodynamics in
one space dimension in a Lorentz invariant way using an instantanous
interaction is closely related to the properties of the linear
potential, as  the interparticle force is constant.
For classical particles, such a constant force
yields the same dynamics as a retarded force \cite{Jackson},
which can be seen in the following way:
The origin of retardation effects is that
no signal can propagate faster than with the speed
of light. Therefore,  the force which acts on a particle at position
$x_M$ (see figure \ref{Abb1}), because of a particle
travelling on the Minkowski worldline $x_T (t) $, is
determined by the intersection point $x_S$ of the
worldline $x_T$ with the backward light cone of $x_M$.
The size and direction of the force only depends on whether the
intersection point $x_{S}$ is right or left of $x_{M}$. Because the particle
cannot move faster than with the speed of light, the intersection point
of the wordline $x_{T}$ with the $t=0$ axis (which determines the
instantaneous force) is on the same side as $x_{S}$ with respect to $x_{M}$.
Therefore, retardation effects are not important for the specific system
considered here.

For our discussion of the quantum Poincar\'e algebra, we
replace the canonical variables $x_n$ and $p_n$ by the
corresponding operators $\hat{x}_{n}$ and $\hat{p}_{n}$.
One finds
that the first two equations in (\ref{Qpoinc}) are fullfilled, but that
\beq
[\hat{K},\hat{H}] = i\hat{P}+\hat{Z},
\eeq
with a nonvanishing operator
\bea
\hat{Z} & := & \frac{1}{2} e^2 \sum_{i=1}^{2}\{\hat{\Omega}_{i}
(\hat{x}_{1}^{2}-\hat{x}_{2}^{2})-(\hat{\Omega}_{i}\hat{x}_{i}+\hat{x}_{i}
\hat{\Omega}_{i})(\hat{x}_{1}-\hat{x}_{2})\}
\label{pl} \\
\hat{\Omega}_{i} & := & [\epsilon 
(\hat{x}_{1}-\hat{x}_{2}),\sqrt{\hat{p}_{i}^{2}
+m^{2}}]. \nonumber
\eea
This means that the Poincar\'e invariance of the classical system is
destroyed by quantization and that we encounter an anomaly.
The anomaly is characterized by the operator $\hat{Z}$.

We will now study matrix elements of this operator in order to
obtain an intuitive understanding of the nature of the anomaly.
In appendix A, it is proven that
\begin{eqnarray}
\langle x_{1}^{'}x_{2}^{'}| \hat{Z}|x_{1}x_{2} \rangle=
 \frac{e^2 m}{2\pi}\{ \epsilon(x_{1}^{'}-x_{2}^{'})-\epsilon(x_{1}-x_{2}) \}
(x_{1}-x_{2})(x_{1}^{'}-x_{2}^{'})\times \nonumber \\
    \{\delta(x_{2}-x_{2}^{'})\frac{K_{1}(m|x_{1}-x_{1}^{'}|)}
{|x_{1}-x_{1}^{'}|}
 -\delta(x_{1}-x_{1}^{'})\frac{K_{1}(m|x_{2}-x_{2}^{'}|)}{|x_{2}-x_{2}^{'}|}\},
\label{ZZZ}
\end{eqnarray}
where $K_{1}$ is a modified bessel function \cite{Abramowitz}.
For the case that the two particles do not cross each other, $(x_{1}-x_{2})$
and $(x_{1}^{'}-x_{2}^{'})$ have the same sign and the  matrix
element vanishes. Therefore, the first conclusion at this point is
that crossing
of the particles must be an essential feature for the appearance of the
anomaly.
We can see this clearly by
calculating
matrix elements of
the first relativistic contribution\footnote{
In the nonrelativistic limit $\sqrt{p^2 + m^2}\approx m + p^2 / (2m)$
the Poincar\'e algebra closes. In this case $K$ is the generator
of Galilei transformations.
}
in (\ref{pl})
\beq
\hat{Z} \approx
\frac{e^2 i}{2m^{3}} \{\delta\xd (\hat{p}_{1} +
\hat{p}_{2}) \}
\label{i11}
\eeq
with eigenfunctions of
the nonrelativistic Hamiltonian
\beq
\hat{H} = \sum_{i=1}^{2}
\frac{\hat{p}_{i}^{2}}{2m} +
e^2 |\hat{x}_{1}-\hat{x}_{2}|.
\eeq
The eigenstates are Airy-functions with the eigenvalues
\beq
E_{n} = |a_{n}^{'}| (\frac{e^{4}}{m})^{\frac{1}{3}} ,
\eeq
where $a_{n}^{'} $ are the zeros of the first derivative of the Airy-function
(see appendix B).
It is easy to show that
$\hat{Z}$ of eq.(\ref{i11}) has the following expectation value
\beq
\langle \Psi_{n}|\hat{Z}|\Psi_{n} \rangle =\frac{ie^4}{4m^{3}}\frac{1}{E_n}
.   \label{konf}
\eeq
Classically, the particles move periodically with period $T$. Because of the
virial theorem, $T \sim \sqrt{E}$, since the potential is
linear.
Therefore eq.(\ref{konf}) can be written
alternatively
 \beq
\langle \Psi_{n}|\hat{Z}|\Psi_{n} \rangle \sim \frac{1}{T^{2}}.
\label{intt}
 \eeq
This result can be explained in the following way:
we already pointed out that  crossing of the particles
is the physical reason for the
anomaly. Therefore  we expect that
the matrix element under consideration is proportional to the
number of crossings in a given time interval, i.e. $\sim 1/T$.
It has also to be proportional to the typical time the particles'
distance is less than one Compton wavelength. This typical time
is proportional to the inverse velocity of the particles when they cross
each other which
is proportional to the inverse oscillation time $1/T$. Therefore
we find that $\langle Z \rangle \sim 1/T^2 $ in agreement with the calculation.

\begin{figure}
%  \epsfxsize=12cm
%  \centerline{\epsffile{fig2.ps}}
\caption{Quantum mechanical particle propagation in Minkowski space.}
\label{Abb2}
\end{figure}
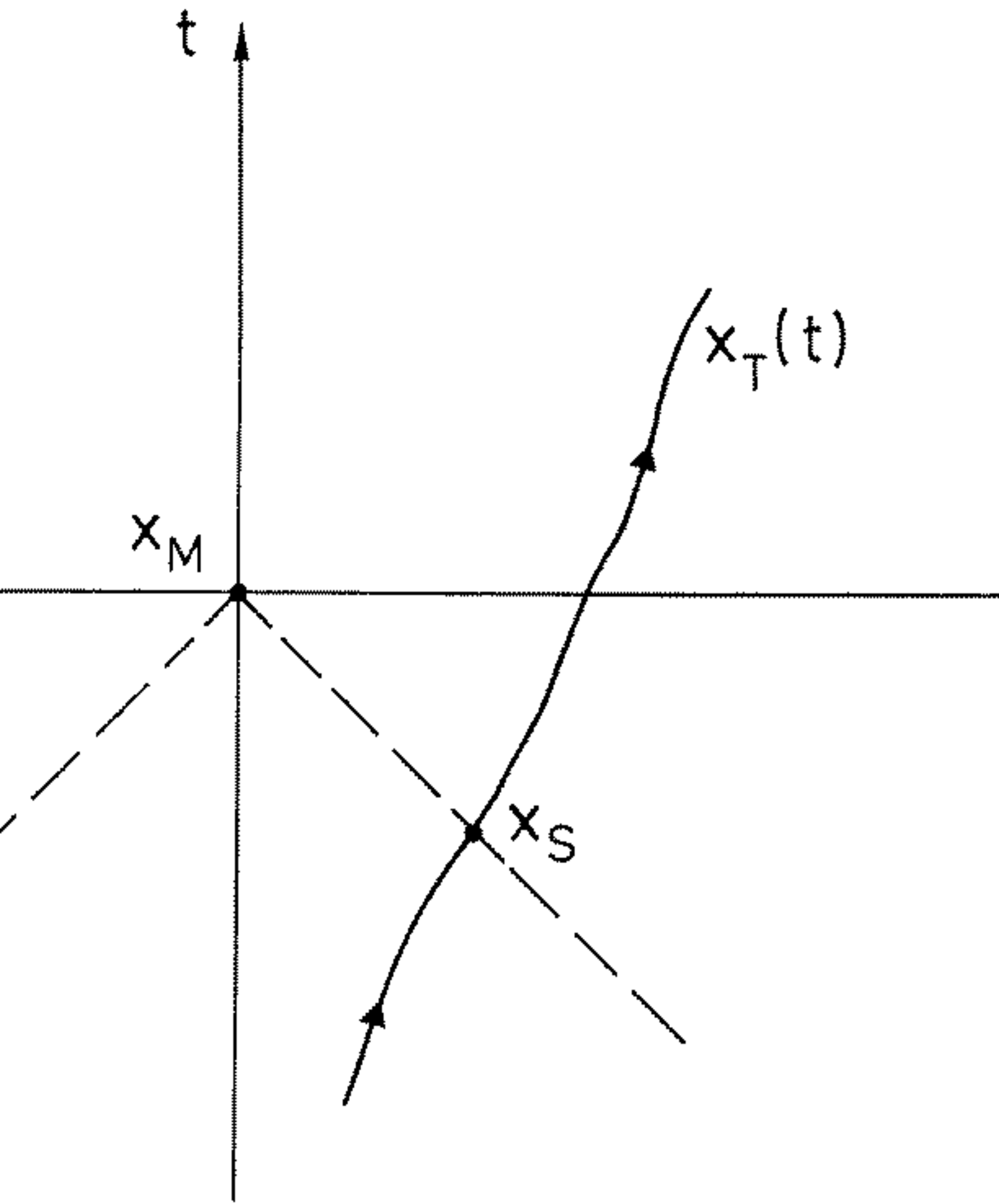
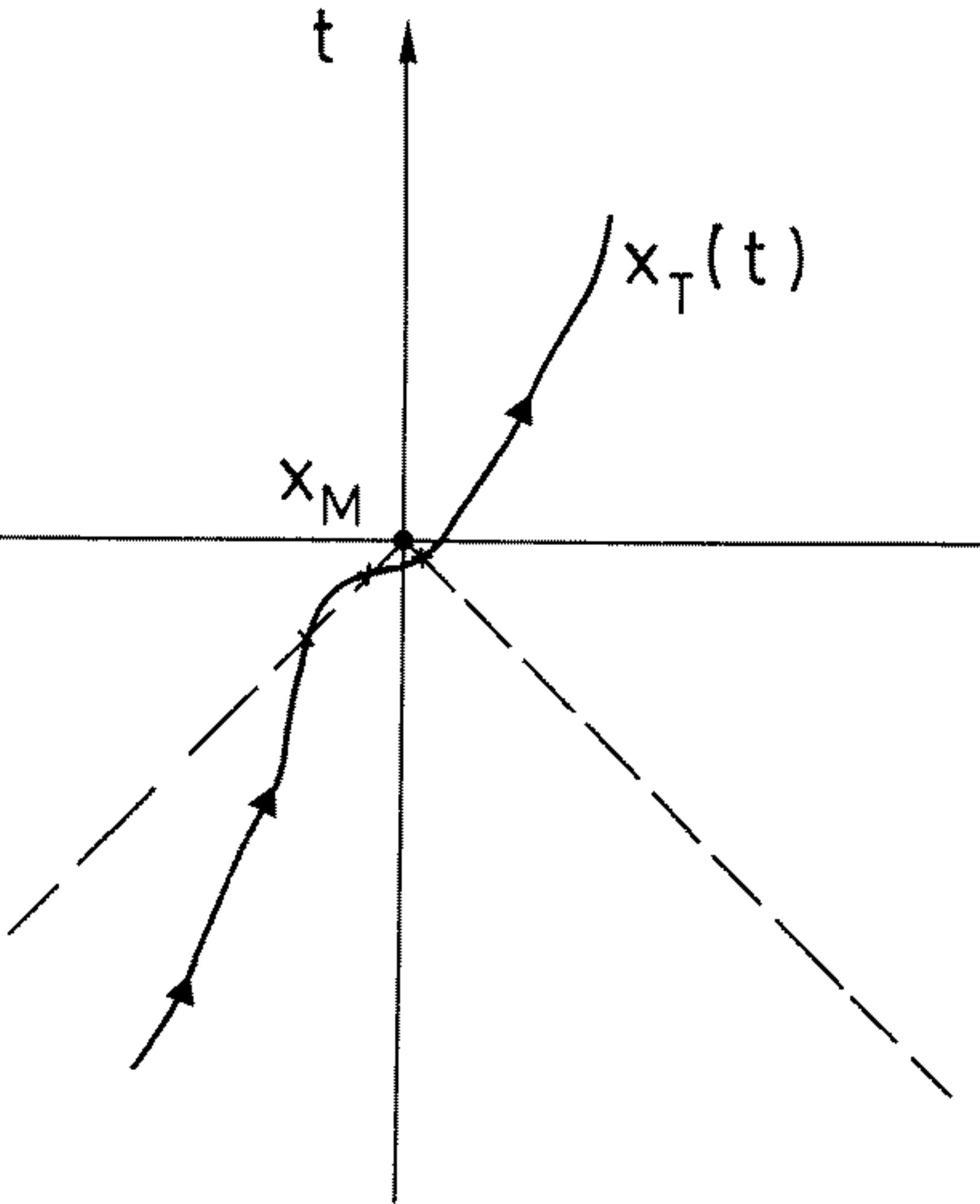

The physical reason for the violation of the Poincar\'e algebra
is related to the fact that quantum mechanically
a particle can move faster than the speed of light over distances
comparable with its Compton wavelength. This one can see by
calculating the free retarded propagator
 of a relativistic particle in 1+1-dimensions
\beq
G_{0}(x',t';x,t)=\theta (t'-t)\frac{m(t'-t)}{\pi
\sqrt{(x'-x)^{2}-(t'-t)^{2}}}K_{1}(m\sqrt{(x'-x)^2-
(t'-t)^2})
\eeq
($\theta (x)$ is 1 for positive and 0 for negative argument).
Propagation into the spacelike region
$(x'-x)^2 > (t'-t)^2 $ is possible although it is exponentially
suppressed as $K_1 (mz)\sim \exp(-mz)$ for large arguments.
Therefore, retardation effects are important if
the particles cross each other. In fig.\ref{Abb2}
there are two intersection
points of the particle worldline with the backward light cone of $x_{M}$
on the left side of $x_{M}$ and one on the right side.
The instantaneous
force in the Hamiltonian formalism does not describe the dynamics
properly because the intersection point of the particle wordline with the
$t=0$-axis is on the right side of $x_{M}$, causing
a negative force. The retarded force, however, is positive.

If a particle
propagates forward in time into a spacelike region in one
Lorentz frame
then there is another Lorentz frame where this particle propagates
backward in time.
This is interpreted as the presence of an antiparticle.
In order to keep
Lorentz invariance one is
forced to allow for particle-antiparticle production
which is not inherent in the quantum mechanical Hamiltonian considered
here.

It is possible to get an approximately closed
Poincar\'e algebra, however, including the first relativistic correction. To 
this
end we add to the Hamiltonian an additional potential
 $h\delta (\hat{x}_{1}-\hat{x}_{2})$ which
is only nonzero when the particles cross each other.
$h$ has to be determined
in such a way that the Poincar\'e algebra is closed up to this
order. We therefore have for the Hamiltonian and the boost generator
\beq
\hat{H}'= \hat{H}+h\delta\xd ,\qquad
\hat{K}'= \hat{K}+\frac{h}{2}(\hat{x}_{1}+\hat{x}_{2})\delta\xd .
\eeq
One finds that the first two equations of (\ref{Qpoinc})
still hold and that
\beq
[\hat{K}',\hat{H}']=i\hat{P}+\hat{Z}+\hat{Z}'
\eeq
where $\hat{Z}$ is defined as in eq.(\ref{pl})
and 
\beq
\hat{Z}'=\frac{hi}{2}\{\ (\frac{\hat{p}_{1}}{\wpa}+\frac{\hat{p}_{2}}{\wpb})
\delta\xd+\delta\xd(\frac{\hat{p}_{1}}{\wpa}+\frac{\hat{p}_{2}}{\wpb})\ \}.
\eeq
With $\wpa \approx m$ and $\wpb \approx m$ we have
\beq
\hat{Z}' \approx \frac{hi}{m}[(\hat{p}_{1}+\hat{p}_{2})\delta\xd].
\eeq
Comparing that with $\hat{Z}$ in eq.(\ref{i11}) we find that
the Poincar\'e algebra is closed including the first relativistic
correction,
if we set
\beq
 h=-\frac{e^2}{2m^{2}} .
\eeq
One obtains an intuitive understanding for this additional attractive
$\delta$-potential by keeping in mind that it is only non-vanishing
when the particles cross each other and it enhances the relative
velocity  during the crossing. This is in agreement with the
qualitative explanation of the anomaly given above.

Higher order relativistic corrections contain derivatives of
the $\delta$-function at the origin. Therefore locality of the
interaction is lost.

\section{Poincar\'e invariance in light-cone
coordinates}
\label{Light}

In the quantized theory
one has the following generators of the Poincar\'e
group
\beq
\hat{H}_{LC}=\sum_{i=1}^{2} \frac{m^{2}}{2\hat{p}_{i}} + e^2
|\hat{x}_{1}-\hat{x}_{2}|
\eeq
\beq
\hat{P}_{LC}=\sum_{i=1}^{2} \hat{p}_i \qquad
\hat{K}_{LC}=\frac{1}{2}\sum_{i=1}^{2} (\hat{p}_{i}\hat{x}_{i} +
\hat{x}_{i}\hat{p}_{i}). \label{lightboost}
\eeq
Note that $\hat{K}$ is independent of the interaction and therefore
a non-dynamical generator.

It is easy to check that the Poincar\'e algebra (\ref{Poinc2})
still closes in the quantum theory.
This result clearly signals that the quantum mechanical system
under consideration must be  different from that in ordinary
coordinates where Poincar\'e invariance was violated by quantization.
The reason for this is rooted in the structural simplicity
of proper Lorentz transformations  on the light-cone.
The Lorentz tensor reads
\beq
\Lambda_{\ \nu}^{\mu}=\left( \begin{array}{cc}
                               \exp(\alpha) & 0 \\
                               0 & \exp(-\alpha)
                             \end{array}
                      \right),
\eeq
where $\alpha$ is the velocity of the moving frame.
As a consequence, Lorentz transformations are simply scale transformations
of the coordinates $x^{+}$ and $x^{-}$ without mixing them
in contrast to ordinary coordinates. Therefore, a particle
which moves forward in time in one Lorentz frame will move forward in
time in any other Lorentz frame independent of the dynamics under
consideration. This fact is reflected in the simple
form of the boost generator in (\ref{lightboost}).

To remove negative energy states
(for which one does not have a proper interpretation in a quantum
mechanical theory), one can introduce
a projection operator on positive momentum states
\beq
\hat{Q}:=\int_{0}^{\infty}dp\ |p \rangle\langle p| .
\eeq
It is easy to check that the projected operators
\beq
\hat{H}_{Q}:=\hat{Q}\hat{H}_{LC}\hat{Q}; \quad
\hat{P}_{Q}:=\hat{Q}\hat{P}_{LC}\hat{Q}; \quad
\hat{K}_{Q}:=\hat{Q}\hat{K}_{LC}\hat{Q}
\eeq
still fulfill a closed Poincar\'e algebra
\beq
[\h_{\q},\p_{\q}]=0, \quad
[\k_{\q},\h_{\q}]=-i\h_{\q}, \quad
[\k_{\q},\p_{\q}]=i\p_{\q}.
\eeq

 For this fact it is important that
the boost operator $\hat{K}_{LC}$ and the projection operator
on positive momentum states $\hat{Q}$
commute ($[\hat{K}_{LC},\hat{Q}]=0$).
This is intuitively clear because of the simple form of the boost
transformation.
 As a consequence, the decomposition of the
Fock space into spaces of definite particle number
and projection on positive momenta is a Lorentz invariant
concept. This phenomenon is also valid in 1+1-dimensional QCD
formulated in light-cone variables where
it can be used for a valence quark approximation of mesons and baryons
which is Lorentz-invariant \cite{Horn}.

\section{The anomaly in the infinite momentum frame}
\label{Infinit}

Weinberg motivated the use of light-cone variables by switching
to a system which moves with infinite momentum \cite{Weinberg}.
We will show now that in this frame the effect of the anomaly
disappears.
Consider the matrix element
$\langle p_{1}'p_{2}'|\hat{Z}|p_{1}p_{2} \rangle$.
After a straightforward, but lengthy calculation one finds
in the infinite momentum limit
$p_{1},p_{2} \gg m\ $ and $p_{1}',p_{2}' \gg m\ $
\beq
\lim_{P\rightarrow\infty}
\langle p_{1}'p_{2}'|\hat{Z}|p_{1}p_{2} \rangle
=
\frac{im^{2}e^2}{4\pi P^{4}}\delta(P-P')(\frac{1}{x^{2}y^{2}}
+\frac{1}{(1-x)^{2}(1-y)^{2}})  \nonumber
\eeq
where
\beq
p_{1}=xP \quad p_{2}=(1-x) P \quad P=p_1 + p_2
\eeq
\beq
p_{1}'=yP \quad p_{2}'=(1-y) P \quad P' = p_1' + p_2' .
\eeq
The anomalous matrix element vanishes in the limit $P\to\infty$.
Therefore it is possible to formulate relativistic quantum mechanics
on a truncated Fock space in the infinite momentum frame.
This is in accordance
with Weinberg's motivation of the light cone.

The anomaly will only vanish if both particles carry a large momentum
fraction, due to the singularities at $x=0$, $y=0$ and $x=1$, $y=1$.
If quantum states with a non vanishing wave function at these
singular points contribute in a relativistic wave equation, then
the effect of antiparticles becomes important.

Moreover, it
is not possible to boost wave functions or non scalar observables
from the infinite momentum frame to another Lorentz frame, as
we have no closed Poincar\'e algebra, which means we have
no generator of unitary transformations $K$ in ordinary
coordinates on a truncated Fock space.
A two-body wave function in the infinite momentum frame
is in general a complicated many-body wave function in the rest frame,
containing a large number of particle-antiparticle pairs.
Two problems arise if one tries to construct this wave function:
first one has to find the boost generator in the presence
of antiparticles, i.e. from the corresponding quantum field theory.
This boost generator will in general be a dynamical quantity
dependent on the interaction under consideration.
The second problem is to calculate the boosted wave function
$\psi'$.
It is given by the equation
\[
 | \psi' \rangle = \exp( i \lambda K ) | \psi \rangle
.
\]
As $K$ is dynamical, i.e. it contains the interaction, this problem
is as complicated as finding the full propagator of the quantum
field theory.
Therefore only scalar
observables can be calculated in the infinite momentum frame
or light-cone coordinates and compared with quantities in the
rest frame of the two particles in ordinary coordinates.

\section{Discussion}
\label{ENDE}

We have seen that classical Poincar\'e symmetry
of electrodynamics in one space dimension is lost if the theory is
quantized on a truncated Fock space in ordinary
coordinates. In a Hamiltonian formulation, 
this shows up as an anomaly in the Poincar\'e algebra.
Almost all known examples of anomalies are restricted to relativistic
quantum field theories. However, the loss of a classical symmetry
after quantization is a more general phenomenon, which is related
to the necessity of specifying a Hilbert space \cite{Raif1}.
This can be seen
in the simple case of the motion of a free particle
in one dimension, with $H=p^2$ \cite{Raif2}. If one quantizes
the theory on the Hilbert space $L_2 [-\infty,\infty]$, the
classical theory and the quantum theory are translationally  invariant.
The momentum operator $p$ generates a unitary transformation
which translates the system according to
\[
   \exp(ip\alpha) x \exp(-ip\alpha) = x + \alpha
  .
\]
If the theory is quantized on a circle with periodic boundary
conditions this symmetry breaks down to a discrete lattice
symmetry, i.e. the parameter $\alpha$ has to be
constrained to integral multiples of $2\pi$ \cite{Raif2}.
For a detailed discussion of a related problem in the presence
of an external  magnetic flux through a circle see \cite{Manton}.
Another example of an anomaly of a finite dimensional system
which is closely related to the axial anomaly in QED, is the
supersymmetric harmonic oscillator. The parity symmetry of
the classical theory is broken by the ground state
but all excited states appear in parity doublets \cite{Raif2}.

The fact that the Poincar\'e anomaly is present in ordinary
coordinates but not in light-cone coordinates indicates
that a different Hilbert space is used in the two formulations.
In light-cone coordinates the Fock space can be decomposed
into spaces of a definite number of particles without
violating Lorentz invariance, whereas in ordinary coordinates
this is not possible. Therefore
it is not possible to compare wave functions in the two
formulations in a simple manner. As in the case of the
infinite momentum frame,
a two-body wave function in light-cone coordinates does
not correspond to a two-body wave function in ordinary coordinates.

The discussion of the anomaly in the infinite momentum frame
exemplifies that our results are consistent with the original
motivation of light-cone coordinates \cite{Weinberg}.
From the structure of the anomalous matrix element in the infinite
momentum frame, one can read off a necessary condition for
the validity of the valence quark approximation in this
Lorentz frame.
If the momentum space wave functions do not vanish rapidly enough at
the singular points $x=0$ and $y=0$, the anomaly is important also
in this frame, therefore the truncation of Fock space violates
Lorentz invariance.

Our results also clearly indicate that the so called ``simplicity''
of light-cone theories is rather deceptive and probably an artefact
of working in 1+1-dimensions, as it relies on the exceptionally
simple structure of the boost generator and the Fock space.
Unfortunately,  it is clear that two of the three boost generators will
become dynamical quantities  in 3+1-dimensions also on the light-cone.
As a result, a Lorentz invariant formulation of quantum mechanics
will be as difficult as in ordinary coordinates.
Therefore the intuition which is gained by studying
lower dimensional theories may be misleading when
phenomena in the real world, i.e. in 3+1-dimensions
are to be understood.
In addition, it will be
much harder to implement e.g. rotational invariance on the light-cone,
because the generators for this symmetry are also dynamical, in contrast to
ordinary coordinates. 
%
%Therefore, the widely spread enthusiasm about the
%power and possibilities of a light-cone formulation of QCD has to
%be taken very sceptically.
%

\section*{Acknowledgements}
We would like to thank M. Burkardt, F. Lenz
and D. Stoll for many fruitful discussions 
and M. Engelhardt and D. Kusnezov for 
reading the manuscript.

\appendix
\section{Calculation of matrix elements of $\hat{Z}$}
In this appendix we compute the matrix element 
$\langle x_{1}^{'}x_{2}^{'}|\hat{\Omega}_{1}|
x_{1}x_{2} \rangle$: 
\begin{eqnarray*}
\lefteqn{\langle x_{1}'x_{2}'|\hat{\Omega}_{1}|x_{1}x_{2} \rangle= } \\
  &= & \langle x_{1}'x_{2}'|\ep \wpa-\wpa \ep|x_{1}x_{2} \rangle =   \\
  &= & \{\epsilon(x_{1}'-x_{2}')-\epsilon(x_{1}-x_{2})\}
      \int_{-\infty}^{\infty}dp_{1}\int_{-\infty}^{\infty}dp_{2}\
      \langle x_{1}'x_{2}'|\wpa|p_{1}p_{2}\rangle
      \langle p_{1}p_{2}|x_{1}x_{2} \rangle =  \\
  &= & \{\ \epsilon(x_{1}'-x_{2}')-\epsilon(x_{1}-x_{2})\ \}
      \delta(x_{2}-x_{2}')\int_{-\infty}^{\infty}dp_{1}
      \sqrt{p_{1}^{2}+m^{2}}\frac{\exp
      (ip_{1}(x_{1}'-x_{1}))}{2\pi}.
\end{eqnarray*}
For the integral on the right side one arrives at 
\begin{eqnarray*}
\lefteqn{\int_{-\infty}^{\infty}\sqrt{p_{1}^{2}+m^{2}}
   \frac{\exp(ip_{1}(x_{1}'-x_{1}))}{2\pi}dp_{1}\ = } \\
  & &=2\int_{0}^{\infty}\sqrt{p_{1}^{2}+m^{2}}\frac{\cos(p_{1}q)}{2\pi}dp_{1}
     =\frac{2}{2\pi}\int_{0}^{\infty}(\sqrt{p_{1}^{2}+m^{2}}-p_{1}+p_{1})
     \cos(p_{1}q)dp_{1}
\end{eqnarray*}
where q is defined as $q:=|x_{1}'-x_{1}|\ $. Define
\beq
I_{1}:=\frac{2}{2\pi}\int_{0}^{\infty}dp_{1}p_{1}\cos(p_{1}q).
\eeq
This integral has to be regularized
\beq
I_{1,reg}:=\frac{2}{2\pi} \lim_{\epsilon\rightarrow 0+}
  {\rm Re\, }\{\int_{0}^{\infty}dp_{1}p_{1}\exp(p_{1}(iq-\epsilon))\}
  =\frac{-1}{\pi q^{2}}.
\eeq
For the integral $I_{2}$ defined as
\beq
I_{2}:=\frac{2}{2\pi}\int_{0}^{\infty}dp_{1}(\sqrt{p_{1}^{2}+m^{2}}-p_{1})
\cos(p_{1}q)
\eeq
one obtains after two partial integrations
\beq
I_{2}=\frac{1}{\pi q^{2}}-\int_{0}^{\infty}dp_{1}\frac{m^{2}
\cos(p_{1}q)}{\pi\sqrt{p_{1}^{2}+m^{2}}^{3}q^{2}}.
\eeq
Therefore one has finally \cite{Abramowitz}
\begin{eqnarray*}
\lefteqn{\int_{-\infty}^{\infty}dp_{1}\frac{\sqrt{p_{1}^{2}+m^{2}}}
{2\pi}\exp(ip_{1}(x_{1}'-x_{1}))= } \\
 & & = I_{1}+I_{2}=-\int_{0}^{\infty}dp_{1}\frac{m^{2}\cos(p_{1}q)}
 {\pi \sqrt{p_{1}^{2}+m^{2}}^{3}q^{2}}=\frac{-m}{\pi
 |x_{1}'-x_{1}|}K_{1}(m|x_{1}'-x_{1}|)
\end{eqnarray*}
Hence
\beq
\langle x_{1}'x_{2}'|\hat{\Omega}_{1}|x_{1}x_{2} \rangle =
\{\ \epsilon(x_{1}'-x_{2}')-\epsilon(x_{1}-x_{2})\ \}
\delta(x_{2}'-x_{2})\frac{-m}{\pi |x_{1}'-x_{1}|}K_{1}(m|x_{1}'-x_{1}|).
\eeq
The calculation for $\langle x_{1}'x_{2}'|\hat{\Omega}_{2}|x_{1}x_{2}\rangle $
is completely analogous. From the two matrix elements (\ref{ZZZ}) follows
immediately.

\section{Solution in the nonrelativistic limit}
The nonrelativistic Hamiltonian for two particles interacting by a linear
potential is
\beq
\hat{H}=\sum_{i=1}^{2}\frac{\hat{p}_{i}^{2}}{2m}+e^2|\hat{x}_{1}-
\hat{x}_{2}|.
\eeq
After separating  the center of momentum one has for the relative variable
the following Schr\"odinger equation
\beq
-\frac{1}{m}\frac{\partial^{2}}{\partial r^{2}}\psi(r)
+e^2|r|\psi(r)=E\psi(r).
\eeq
For $r>0\ $ one gets after the substitutions  $y:=(e^2r-E)m\ $ and
$z:= (e^{4}m^{2})^{-\frac{1}{3}}y\ $
\beq
\frac{d^{2}\psi(z)}{dz^{2}}-z\psi(z)=0.
\eeq
The solution is \cite{Abramowitz}
\beq
\psi(z)=c_{1} {\rm Ai}(z)+c_{2}{\rm Bi}(z).
\eeq
$c_{2}$ must be zero because ${\rm Bi}(z)$ is not normalizable.
For $r<0$ the procedure is similar.
For the wave function one gets then
\beq
\psi(r)\sim Ai(m(e^{4}m^{2})^{-\frac{1}{3}}(e^2|r|-E)).
\eeq
The energy eigenvalues are determined by the requirement that the wavefunction
and its first derivative is continuous at the origin. Therefore one has
\beq
Ai'(-m(e^{4}m^{2})^{-\frac{1}{3}}E)=0
\eeq
and if the zeros of the first derivative of the Airy-function are denoted
$a_{n}'$ one has finally
\beq
E_{n}=|a_{n}'|(\frac{e^{4}}{m})^{\frac{1}{3}}.
\eeq
$|\psi_{n}(0)|^{2}$ is given by
\beq
|\psi_{n}(0)|^{2}=\frac{1}{|a_{n}'|}(\frac{e^2 m}{8})^{\frac{1}{3}}
\eeq
and therefore we have for $\langle \psi_{n}|\hat{Z}|\psi_{n}\rangle$
\beq
\langle \psi_{n}|\hat{Z}|\psi_{n}\rangle =\int dr\, \psi(r)^{*}\frac{e^2 i}
{2m^{3}}\delta(r)\psi(r)=\frac{ie^2}{2m^{3}|a_{n}'|}
(\frac{e^2 m}{8})^{\frac{1}{3}}.
\eeq

\newpage
%
% uncomment the commented lines to include the figures:
\begin{figure}
%  \epsfxsize=12cm
%  \centerline{\epsffile{fig1.ps}}
\caption{Classical particle trajectory in Minkowski space.}
\label{Abb1}
\end{figure}
%\newpage
\end{document}